\documentclass[10pt,twocolumn,amsmath,amssymb,floatfix]{revtex4}
\usepackage{graphicx,epsfig,latexsym,amssymb}
\usepackage{multirow,amsmath,array,booktabs}
\usepackage{dcolumn}
\usepackage{color}
\usepackage{textcomp}
\usepackage{longtable}
\usepackage[colorlinks,linkcolor=blue,citecolor=blue]{hyperref}
\usepackage[section]{placeins}
\usepackage{bm}

\begin{document}

\title{One neutron triaxial halo candidates in aluminum isotopes from reaction observables}

\author{Jia-Lin An$^{1,3}$}
\author{Shi-Sheng Zhang$^{1}$}\email{zss76@buaa.edu.cn}
\author{Kaiyuan Zhang$^{2}$}\email{zhangky@caep.cn}

\affiliation{$^{1}$School of Physics, Beihang University, Beijing, 100191, China}
\affiliation{$^{2}$National Key Laboratory of Neutron Science and Technology, Institute of Nuclear Physics and Chemistry, China Academy of Engineering Physics, Mianyang, Sichuan, 621900, China}
\affiliation{$^{3}$Baoding Hospital of Beijing Children's Hospital, Baoding, Hebei, 071000, China}

\begin{abstract}

\noindent\textbf{Abstract: Microscopic description of one neutron ($1n$) halo candidates $^{40,42}$Al, with particular triaxial shape, is presented by combining the triaxial relativistic Hartree-Bogoliubov theory in continuum (TRHBc) with the Glauber reaction model for the first time.
In this scheme, the reaction cross sections of aluminum isotopes on a carbon target at 240 and 900 MeV/A are calculated, which exhibit a pronounced increase for $^{40,42}$Al + $^{12}$C deviating from the systematic trend of their neighbours.
Furthermore, the predicted longitudinal momentum distributions of the residues after $1n$ removal reactions for $^{40,42}$Al + $^{12}$C are narrower than those for $^{36,38}$Al + $^{12}$C, which suggest halo structure with spatially extended density distribution.
Based on the large occupation probabilities of $p$-wave valence neutrons, we identify $^{40,42}$Al as the first triaxially deformed $1n$ $p$-wave halo candidates.
This work cast a new light on the search for the heavier halo nuclei for future experiments in the mass region of $A\approx40$, through theoretical predictions from triaxial structure to reaction observables.}

\end{abstract}

\maketitle

\newpage
\section{Introduction}\label{introduction}

The discovery of the halo nucleus $^{11}$Li with two dilute distributed valence neutrons in a weekly bound system~\cite{Tanihata1985PRL55.2676} started the era of a novel topic in nuclear physics. 
From then on, great efforts have been made to identify halo nuclei from experiments and study the mechanism of the halo formation theoretically. 
So far, less than 20 halo nuclei have been confirmed and the heaviest one is still $^{37}$Mg~\cite{Kobayashi2014PRL_242501,takechi2014evidence}. To search for new heavier halo candidates in the mass region of $A\approx40$ is strongly desired as a breakthrough for future experiments.

Currently, the description from microscopic structure to reaction observables has been realized via the deformed relativistic Hartree-Bogoliubov theory in continuum (DRHBc)~\cite{Zhou2010PRC,Li2012PRC,Zhang2020PRC,Pan2022PRC} combined with the Glauber model. 
This approach has been successfully applied to describe one neutron ($1n$) $p$-wave halo nuclei both $^{31}$Ne~\cite{Zhong2022SCPMA,Cong2024PLB} and $^{37}$Mg~\cite{ZhangKY2023PLB,AnJL2024PLB} with assumption of axial deformations, superior to other approaches, such as asymmetric molecular dynamics (AMD) model with resonating group method (RGM)~\cite{Minomo2012PRL,watanabe2014}, etc. 
Later on, Lu studied the soft electric dipole response of deformed halo nuclei $^{31}$Ne and $^{37}$Mg using a deformed Woods-Saxon potential~\cite{lu2025PRC}.
Lately, the triaxial relativistic Hartree-Bogoliubov theory in continuum (TRHBc) has been developed~\cite{Zhang2023PRC,Zhang2025PRC}, which inherits the merits of the DRHBc theory and further incorporates the triaxial deformation degrees of freedom. 
It can reproduce the location of the proton drip-line, one-neutron separation energies ($S_{n}$), and charge radii for aluminum isotopes~\cite{Zhang2023PRC,Zhang2025PRC}. 
More importantly, the TRHBc theory predicted that the last bound odd-neutron aluminum isotope $^{42}$Al possesses a triaxially deformed halo structure~\cite{Zhang2023PRC, xiang2023Symmetry}. 
However, the predicted radii and the density distributions of neutron-rich aluminum isotopes can not be directly measured from experiments. 
Therefore, the reaction observables for triaxially deformed aluminum isotopes, especially for $^{42}$Al are desirable via a reaction model based on the structure information from the TRHBc theory.

In this work, we adopt the density of the core nucleus and the wave function of the valence nucleon from the TRHBc theory as the inputs of the Glauber reaction model to calculate the reaction cross sections (RCSs) of aluminum isotopes on a carbon target, and the longitudinal momentum distributions of the residues after $1n$ removal reactions for $^{36,38,40,42}$Al + $^{12}$C. 
We aim at providing the quantitative evidence for the triaxial halos from neutron-rich aluminum isotopes for future experiments.

This article is made up of four parts as follows. 
In Section II, we briefly review the theoretical framework of the TRHBc + Glauber approach. 
Then, we show the predictions of reaction observables with this approach in Section III.
Finally, we make a summary in Section IV.

\section{Theoretical Framework}

The structural inputs for the Glauber reaction model, including the angle-averaged densities of the core nucleus and the wave function of the valence neutron, are extracted from the TRHBc theory.
A detailed description of the TRHBc theory can be found in Refs.~\cite{Zhang2023PRC,Zhang2025PRC}.
The Woods-Saxon basis of Dirac equations ~\cite{Zhou2003PRC,Zhang2022PRC} is obtained in a box of 20 fm with the step of 0.1 fm.

Those core nuclear densities $\rho_C(q)$ and the wave functions of valence neutron $\varphi_0$ are put into the Glauber reaction model~\cite{abu2003cross, glauber1959lectures, Horiuchi2010PRC, Fang2004PRC, MaZY2001} to calculate the reaction observables of neutron-rich aluminum isotopes on a carbon target. 
Via this combination, the RCSs between a projectile (P) nucleus and target nucleus (T) can be calculated by the expression
\begin{equation}
    \sigma_{R}(P+T) = \int {\rm d} \bm{b}(1-| \langle{\varphi_0}|e^{i\chi_{CT}(\bm{b}_C)+i\chi_{NT}(\bm{b}_N)}|{\varphi_0}\rangle|^2).
\end{equation}

Another observable, the longitudinal momentum distribution of the fragments after break-up reaction, can be derived and written as~\cite{abu2003cross} 
\begin{align}
\label{momentum distribution}
\nonumber&\frac{{\rm d}\sigma_{-N}^{inel} }{{\rm d}\bm{p}_{\parallel} } 
=\frac{1}{2\pi\hbar}\times\\
&\int {\rm d}\bm{b}_N\left[1-e^{-2Im\chi_{NT}(\bm{b}_N)}\right]\int {\rm d}\bm{s}e^{-2Im\chi_{CT}(\bm{b}_C)}\times\\ 
\nonumber&\int {\rm d}z\int {\rm d}{z'}e^{\frac{i}{\hbar}\bm{p}_{\parallel}(z-{z'})}u_{nlj}^*(r')u_{nlj}(r)\frac{1}{4\pi}P_l({\hat{\bm{r}}}\cdot{\hat{\bm{r'}}}).
\end{align}

It can be seen that these observables are mainly determined by the phase-shift function $\chi$, which reflects the information of the interactions. The phase shift function ~$\chi_{CT}$ ($\chi_{NT}$) denotes the interaction between core (whole) nucleus and the target, and reads
\begin{gather}
    i\chi_{CT} = \int q\rho_C(q) \rho_T(q) f_{NN}(q) J_0(qb) {\rm d} q, \\
    i\chi_{NT} = \int q\rho_T(q) f_{NN}(q) J_0(qb) {\rm d} q.
\end{gather}
 
Via Fourier transform, the phase-shift function in spherical Glauber code can be obtained with the density of the core / target nucleus directly from the nuclear structure models.
Details of the Glauber reaction model can be found in Refs.~\cite{Zhong2022JPG, Zhong2022SCPMA, AnJL2024PLB, wang2024EPJA, An2025JPG} and the references therein.

\section{Reaction observables for neutron-rich aluminum isotopes on a carbon target}

To search for the measurable evidence for triaxial halo nuclei in neutron-rich aluminum isotopes, we calculate the reaction observables using the Glauber model, with the inputs - the densities for the core nucleus and the wave function for valence neutron - from the TRHBc calculations employing the density functional NLSH~\cite{SHARMA1993PLB}.

\begin{figure}[htbp]
  \centering
  \includegraphics[width=0.85\linewidth]{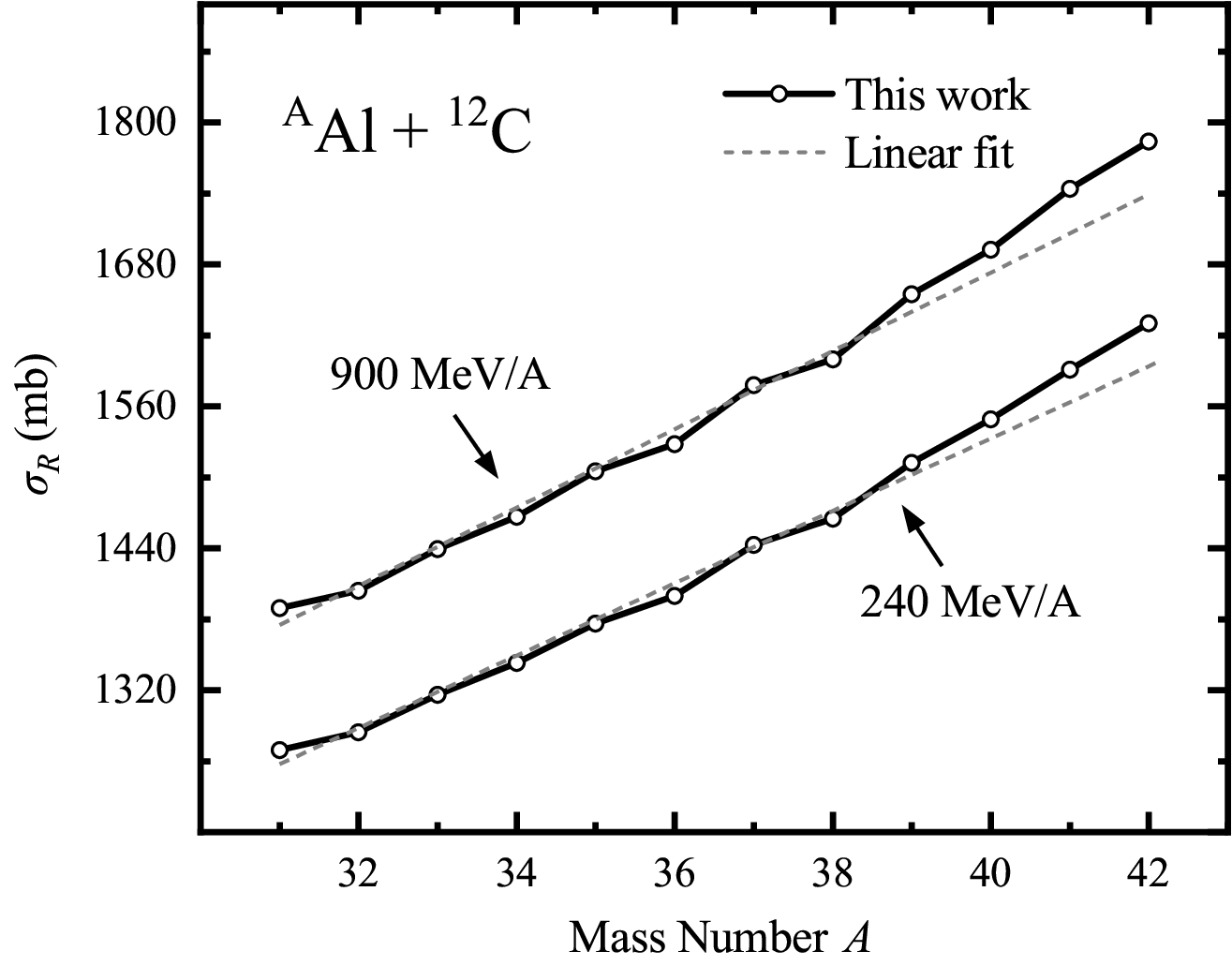}
  \caption{(Color online) RCSs $\sigma_R$ of aluminum isotopes on a carbon target at 240 and 900 MeV/A, respectively. 
   The open black circles denote the cross sections calculated with the Glauber model with inputs from the TRHBc theory. 
   The gray dashed line represents a fit to the cross sections of $^{31-39}$Al + $^{12}$C.
  }
  \label{fig:sigma}
\end{figure}

As one of the direct reaction observables, the RCSs of neutron-rich aluminum isotopes $^{31-42}$Al on a carbon target are displayed in Fig.\ref{fig:sigma}. 
The calculations were performed at incident energies of 240 and 900 MeV/A, respectively. 
The gray dashed line in the plot represents a fit to the RCSs of $^{31-39}$Al + $^{12}$C. 
It can be seen from Fig.\ref{fig:sigma} that there is a pronounced deviation from this trend (gray dashed line) for the RCSs of $^{40,42}$Al on a carbon target at both incident energies. 
Such an increase of the RCSs indicates spatial extension of the density distribution, which can be regarded as a hint of halo formation.

\begin{figure}[htbp]\centering
\includegraphics[width=0.85\linewidth]{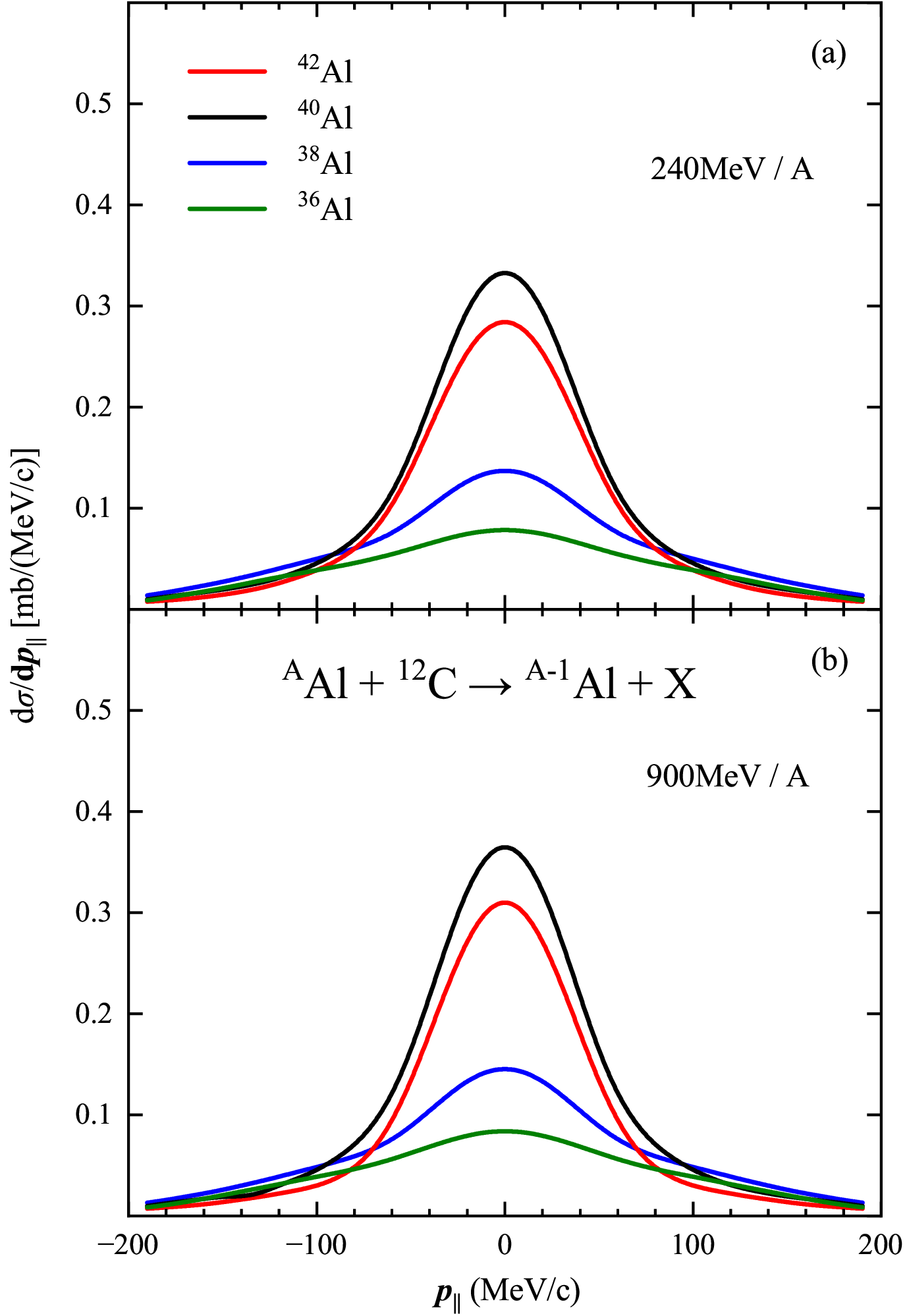}
\caption{(Color online) (a) Inclusive longitudinal momentum distributions d$\sigma/$d$\bm{p}_{\parallel}$ [mb/(MeV/c)] of residues after the break-up reactions $^{36,38,40,42}$Al + $^{12}$C at the incident energies (a) 240~MeV/A, (b) 900~MeV/A, respectively. 
The red/black/blue/green solid line refers to the predictions for $^{42}$Al/$^{40}$Al/$^{38}$Al/$^{36}$Al + $^{12}$C using the TRHBc + Glauber approach.}
\label{fig:momdist}
\end{figure}

Then, we plot the longitudinal momentum distributions d$\sigma/$d$\bm{p}_{\parallel}$ of the fragments after $1n$ removal reactions of $^{36,38,40,42}$Al bombarding a carbon target at 240 MeV/A and 900 MeV/A in Fig.~\ref{fig:momdist} (a) and Fig.~\ref{fig:momdist} (b), respectively.
It can be seen that the momentum distributions of the residues after the break-up reactions $^{36,38}$Al + $^{12}$C are relatively flat, consistent with the picture of deeply-bound valence neutrons structure confined within the nucleus.
In contrast, the momentum distributions for the residues after the break-up reactions $^{40,42}$Al + $^{12}$C are significantly narrower than those for $^{36,38}$Al + $^{12}$C, which suggest a halo structure with dilute distribution of the valence neutrons.
Quantitatively, the full widths at half maximum for $^{40,42}$Al in Fig.~\ref{fig:momdist} are approximately $30\%$ and $50\%$ smaller than those for $^{38}$Al and $^{36}$Al, respectively. 

\begin{table}[htbp]
\small
\caption{Main components and occupation probabilities of the valence neutron orbital in $^{36,38,40,42}$Al (Normalized)}\label{tb:components}\centering\renewcommand{\arraystretch}{1.5}
\begin{tabular}{lrrrr} \toprule[0.8pt]
\multirow{2}{*}{Nucleus} & \multicolumn{4}{c}{Main components}     \\ \cmidrule[0.5pt]{2-5}
                       &$2p_{1/2}$~~~&$2p_{3/2}$~~~&$1f_{5/2}$~~~&$1f_{7/2}$~~~\\ \midrule[0.8pt]
$^{36}$Al              &    $-$~~~~   &    7.5\%   &      3.4\%  &    89.1\%     \\
$^{38}$Al &     3.6\%   &    13.2\%   &     5.9\%  &   77.3\%     \\
$^{40}$Al            &    38.3\%   &    20.3\%   &      7.2\%  &   34.2\%      \\
$^{42}$Al                &    43.2\%  &    13.4\%  &     7.9\%  &  35.5\%      \\ 
\bottomrule[0.8pt]
\end{tabular}
\end{table}

To clarify the contributions from the configurations, we list the components of the valence neutron orbital and their occupation probabilities in Table \ref{tb:components}.
One can see that the contributions from $p$-wave components are nearly $60\%$ in $^{40,42}$Al, whereas those from $f$-wave components are dominated in $^{36,38}$Al.
Therefore, the dominated $p$-wave components play a crucial role on narrow momentum distributions.

\section{Summary}

In all, we search for heavier halo candidates with specially triaxial shape in the mass region of $A\approx40$.
Via the TRHBc + Glauber approach, we study the reaction observables of neutron-rich aluminum isotopes on a carbon target for the first time. 

In this scheme, the RCSs show a significant and energy-independent enhancement for $^{40,42}$Al + $^{12}$C compared to their neighbors, which indicates a clear deviation from the conventional scaling trend.
Furthermore, we study the longitudinal momentum distributions of the residues after $1n$ removal reactions of $^{36,38,40,42}$Al on a carbon target at intermediate and high incident energies.
It turns out that the momentum distributions of the residues after $^{40,42}$Al + $^{12}$C are considerably narrower than those of $^{36,38}$Al + $^{12}$C, in agreement with the extended density distribution according to the Heisenberg uncertainty principle. 
This narrow momentum distribution can be self-consistently interpreted as the dominant $p$-wave components of the valence neutron.

Via a description from the microscopic structure to reaction observables, we have the confidence to conclude that $^{40,42}$Al are promising candidates for the first $1n$ $p$-wave triaxial halo nuclei. In such a way, we cast a new light on the search and identification of the heavier halo nuclei in the mass region of $A\approx40$ at next-generation radioactive beam facilities.
~\\
\section*{ACKNOWLEDGEMENTS}

This work was supported by the National Natural Science Foundation of China (Grant Nos.~12575122, 12175010, 12305125) and the National Key Laboratory of Neutron Science and Technology (Grant No. NST202401016). 

\normalsize \vskip0.3in\parskip=0mm \baselineskip 18pt

\bibliographystyle{iopart-num}
\bibliography{40_42Al_12C}

\end{document}